\documentclass[11pt]{article}
\usepackage{latexsym}
\usepackage{psfrag}
\usepackage[bf,footnotesize]{caption}   
\usepackage[dvips]{epsfig}
\usepackage{amssymb}
\usepackage{graphicx}
\usepackage{graphics}
\begin{document}
\setlength{\unitlength}{1mm}


\newcommand{\ket}[1] {\mbox{$ \vert #1 \rangle $}}
\newcommand{\bra}[1] {\mbox{$ \langle #1 \vert $}}
\def\vac{\ket{0}} \def\vacin{\ket{0}_{in}} \def\vacout{\ket{0}_{out}}
\def\thermal{\ket{\beta}}
\def\bvac{\bra{0}}\def\bvacin{{}_{in}\bra{0}}\def\bvacout{{}_{out}\bra{0}}
\def\bthermal{\bra{\beta}}
\newcommand{\ave}[1] {\mbox{$ \langle #1 \rangle $}}
\newcommand{\avew}[1] {\mbox{$ \langle #1 \rangle $}_w}
\newcommand{\vacave}[1] {\mbox{$ \bvac #1 \vac $}}
\newcommand{\thermalave}[1] {\mbox{$ \bthermal #1 \thermal $}}
\newcommand{\scal}[2]{\mbox{$ \langle #1 \vert #2 \rangle $}}
\newcommand{\expect}[3] {\mbox{$ \bra{#1} #2 \ket{#3} $}}
\def\a{\hat{a}}\def\A{\hat{A}}
\def\b{\hat{b}}\def\B{\hat{B}}
\def\aa{\tilde{a}}\def\AA{\tilde{A}}
\def\bb{\tilde{b}}\def\BB{\tilde{B}}
\def\kplus{\ket{+}}\def\kmoins{\ket{-}}
\def\bplus{\bra{+}}\def\bmoins{\bra{-}}
\def\om{\omega}
\def\ka{\kappa}

\def\p{\prime}
\def\t{\tau}
\def\om{\omega}\def\Om{\Omega}
\def\ga{\gamma}
\def\omp{\om^\p}\def\Omp{\Om^\p}
\def\la{\lambda}\def\lap{\lambda^\p}
\def\mup{\mu^\p}\def\lp{l^\p}
\def\kp{k^\p}\def\sig{\sigma}
\def\ka{\kappa}
\def\al{\alpha}\def\alb{\bar\alpha}
\def\bt{\beta}\def\btb{\bar\beta}
\def\e{\epsilon}
\def\psip{\stackrel{.}{\psi}}\def\fp{\stackrel{.}{f}}
\def\lab{{\bar \la}}
\def\ffi{\varphi}
\def\scry{{\cal J}}
\def\scryp{{\cal J}^+}
\def\scrym{{\cal J}^-}
\def\scrypL{{\cal J}^+_L}
\def\scrymR{{\cal J}^-_R}
\def\scrypR{{\cal J}^+_R}
\def\scrymL{{\cal J}^-_L}
\def\Pig{{\mathbf \Pi}}

\def\di{\partial}
\def\diU{\di_U}
\def\diUU{\di_{\bar U}}
\def\diV{\di_V}
\def\diVV{\di_{\bar V}}
\def\didiv{\raise 0.1mm \hbox{$\stackrel{\leftrightarrow}{\di_V}$}}
\newcommand{\didi}[1]{\raise 0.1mm \hbox{$\stackrel{\leftrightarrow}{\di_{#1}}$}}

\def\d{\mbox{d}}


\def\TUU{T_{UU}}
\def\TUUI{T_{UU}^I}\def\TUUII{T_{UU}^{II}}
\def\Tuu{T_{uu}}
\def\TVV{T_{VV}}
\def\TVVI{T_{VV}^I}\def\TVVII{T_{VV}^{II}}
\def\Tvv{T_{vv}}
\def\re{\mbox{Re}}
\def\im{\mbox{Im}}
\def\S{{\mathbf S}}
\def\T{{\mathbf T}}
\def\1{{\mathbf 1}}
\def\Ss{\mbox{$\hat S$}}
\def\f{{\tilde f}}
\def\ubl{\bar u_L}
\def\UU{{\bar U}}
\def\VV{{\bar V}}
\def\LL{{\cal L_{\rm int}}}
\def\disp{\displaystyle}
\def\bitem{\begin{itemize}}
\def\eitem{\end{itemize}}
\def\bes{\begin{description}}
\def\es{\end{description}}
\newcommand{\be} {\begin{equation}}
\newcommand{\ee} {\end{equation}}
\newcommand{\ba} {\begin{eqnarray}}
\newcommand{\ea} {\end{eqnarray}}
\newcommand{\bsub}{\begin{subeqnarray}}
\newcommand{\esub}{\end{subeqnarray}}

\def\cf{{\it cf}~}
\def\ie{{\it i.e.}~}
\def\etc{{\it etc}}
\def\eg{{\it e.g.}~}
\def\apriori{{\it a priori}~}

\newcommand\dddot[1]{\stackrel{...}{#1}}

\def\nn{\nonumber \\}
\newcommand{\reff}[1]{Eq.(\ref{#1})}

\newcommand{\encadre}[1]{\begin{tabular}{|c|}\hline{#1}\\ \hline  \end{tabular}}

\def\arccosh{\mbox{arccosh}}
\def\l{\left}
\def\r{\right}
\def\Si{\mbox{Si}}
\def\half{{1 \over 2}}
\newcommand{\inv}[1]{\frac{1}{#1}}
\def\inte{\int_{-\infty}^{+\infty}}
\def\into{\int_{0}^{\infty}}
\newcommand\Ie[1]{\inte \! \mbox{d} #1 \;}
\newcommand\Io[1]{\into \! \mbox{d} #1 \;}
\newcommand\IeIe[2]{\int \!\!\! \inte \! \mbox{d} #1 \: \mbox{d} #2 \;}
\newcommand\IoIo[2]{\int \!\!\! \into \! \mbox{d} #1 \: \mbox{d} #2 \;}
\newcommand\Iomoins[1]{\int_{-\infty}^{0} \! \mbox{d} #1 \;}
\newcommand\Iinfsup[3]{\int_{#1}^{#2} \! \mbox{d} #3 \;}
\newcommand{\erf}{\mathop{\rm erf}\nolimits}

\overfullrule=0pt \def\sqr#1#2{{\vcenter{\vbox{\hrule height.#2pt
          \hbox{\vrule width.#2pt height#1pt \kern#1pt
           \vrule width.#2pt}
           \hrule height.#2pt}}}}
\newcommand\lrpartial[1]{\mathrel{\partial_{#1}\kern-1em\raise1.75ex\hbox{$\leftrightarrow$}}}

\newcommand{\doublesum}[4]{\sum_{\begin{array}{cc}\scriptstyle #2 \\ \scriptstyle #1
\end{array}}^{\begin{array}{cc}\scriptstyle #4 \\ \scriptstyle #3 \end{array}}}

\newcommand{\lafootnote}{\fnsymbol{footnote}}

\newcounter{subequation}[equation] \makeatletter
\expandafter\let\expandafter\reset@font\csname reset@font\endcsname
\newenvironment{subeqnarray}
  {\arraycolsep1pt
    \def\@eqnnum\stepcounter##1{\stepcounter{subequation}{\reset@font\rm
      (\theequation\alph{subequation})}}\eqnarray}%
  {\endeqnarray\stepcounter{equation}}
\makeatother

\newcommand{\exergue}[2]{\begin{flushright}{\it\scriptsize #1} {\bf\scriptsize #2}\end{flushright}}


\begin{titlepage}
\hbox to\hsize{%

  \vbox{%
        \hbox{WIS-Detector}%
        \hbox{\today}%
       }}

\vspace{3 cm}

\begin{center}
\Large{Unruh Effect for General Trajectories}

\vskip10mm
\large
{N. Obadia and M. Milgrom}

\vspace{1 cm}

{\small\sl
Center for Astrophysics\\
The Weizmann Institute of Science\\
 Rehovot, Israel\\}
\end{center}

\vspace{20 mm}
\begingroup \addtolength{\leftskip}{1cm} \addtolength{\rightskip}{1cm}

\begin{abstract}

We consider two-level detectors coupled to a scalar field and
moving on arbitrary trajectories in Minkowski space-time. We first
derive a generic expression for the response function using a
(novel) regularization procedure based on the Feynmann
prescription that is explicitly causal, and we compare it to other
expressions used in the literature. We then use this expression to
study, analytically and numerically, the time dependence of the
response function in various non-stationarity situations. We show
that, generically, the response function decreases like a power in
the detector's level spacing, $E$, for high $E$. It is only for
stationary world-lines that the response function decays faster
than any power-law, in keeping with the known exponential behavior
for some stationary cases. Under some conditions the
(time dependent) response function for a non-stationary world-line
is well approximated by the value of the response function for a
stationary world-line having the same instantaneous acceleration,
torsion, and hyper-torsion. While we cannot offer general
conditions for this to apply, we discuss special cases; in
particular, the low energy limit for linear space trajectories.

\end{abstract}

\endgroup
\end{titlepage}
\newpage

\section{Introduction}

The physics of accelerated quantum systems in Minkowski space-times
share some interesting features with quantum gravity phenomena that
are much less tractable; e.g. with the Hawking evaporation of black
holes\cite{Hawking} and quantum aspects of expanding universes and
inflation\cite{Inflation}. Beside being of great interest on their
own right, such Minkowskian phenomena may shed light on their full
fledged curved space time cousins.

A moving two-level detector (so called a Unruh or de Witt detector)
that couples to the modes of a field of interest is a simple and
very useful paradigm for studying quantum effects of accelerated
systems. The most renowned of the phenomena studied with such a
device is the Unruh effect\cite{Unruh}, whereby a linear, uniformly
accelerated detector in flat space-time perceives the vacuum as a
thermal bath of the modes to which it couples. (Whenever we refer in
the paper to space characteristics of a world-line, such as its
being linear, circular, \etc, we mean it to be so in some Lorentz
frame.) One starts by defining a vacuum--the ground state of the
various fields of interest--based on a set of creation-annihilation
operators associated with inertial observers. This means that modes
of any free field are eigenmodes with respect to the proper-time of
such inertial observers.

The vacuum effects associated with the detector are encapsuled in
the response function, which is the object we study in this paper.
The correct way to regularize the divergent expression for the
two-point Wightman function, which appears in the definition of the
response function, is still a moot issue. For example, following
work by Takagi\cite{Takagi}, Schlicht\cite{Schlicht} proposed to
start with a detector of finite spatial width in the proper frame.
In this case, the Wightman function is regular, and the point
detector response function is gotten by taking the width to zero.
This gives well-defined expressions, which have been generalized by
others \cite{Langlois,LoukoSatz}. We discuss the merits of this
approach, and propose an alternative that is, in our view, superior
in some regards.

The response function is an explicit function of $\t$, the
proper-time along the world-line, and of $E$; it is also a
functional of the world-line. Since the response function is
Poincar\'e invariant, it is reduced, in $3+1$ dimensions, to a
functional of the three world-line invariants: the acceleration
$a(\t)$, the torsion $T(\t),$ and the hyper-torsion $H(\t)$.
Beside the trivial, inertial case, only two world-lines have so
far lent themselves to analytic evaluation of the response
function: the original, Unruh case (constant $a$, $T=H=0$) , and
planar cusp motion (constant $a=T$, $H=0$), which is also the
infinite-Lorentz-factor limit of uniform, circular motion response
function, as we show below. The high-$\gamma,$ circular Unruh
effect has been discussed extensively (e.g., by Bell and
Leinaas\cite{BellLeinaas}, and by Unruh\cite{UnruhCircular}), who
obtained analytic results for the infinite Lorentz factor case, in
the hope that it can be detected in electrons moving in storage
rings. Letaw and Pfautsch\cite{LetawPfautschCirc} considered,
numerically, circular motion with finite velocities. Experimental
manifestations of the linear Unruh effect have been discussed in
\cite{ChenTajima,Habs}. The above are all examples of stationary
world-lines, for which the response function is constant. Our main
goal is to get some insight into properties of the response
function for more general, non-stationary world-lines.

All along we assume that the interaction of the detector with the
relevant field is ever present and constant with time. Otherwise the
response function will be affected. For example even inertial
detectors will be characterized by a non-zero response function if
the interaction strength is variable.

The metric is $(+,-,-,-)$ and, except when expressed explicitly, the
value of the constants $(c,\hbar,G,k_B)$ is taken as unity.

Section 2 is dedicated to the discussion of some general
properties of the response function including our proposed
regularization. This is then applied in section 3 to derive
various numerical and analytical results for non-stationary
trajectories. For example, we give integral representations of the
transition rate as function of the ratios $E/a$ and $E/T$ in $1+1$
and $2+1$ dimensions. We also constrain the asymptotic behavior of
the response function in the high $E$ regime. Section 4 summarizes
our conclusions.

\section{The response function}

\subsection{Preliminaries}
We start with some well documented preliminaries. One considers a
massless, chargeless scalar field $\Phi(t,\vec{x})$. In a flat
Minkowski space-time one can use particle states associated with
inertial observers and decompose the field operator into these
states according to the standard definition: \ba\label{Phi}
\Phi(t,\vec{x}) = \int \!\!\!\!\!\int \!\!\!\!\! \inte \!
\frac{\mbox{d}\vec{k}}{\sqrt{2{(2\pi)}^3||\vec{k}||}} \; \left(
a_{\vec{k}} \, e^{-i k_\mu  x^\mu} \ + \ a_{\vec{k}}^\dagger \,
e^{+i k_\mu  x^\mu} \right) \ , \ea where the annihilation and
creation operators satisfy the usual commutation rules
$[a_{\vec{k}},a_{\vec{k'}}^\dagger]=\delta(\vec{k}-\vec{k'})$. We
couple this field to a point-like moving detector constrained to
follow the time-like world-line $x^\mu(\t)=[t(\t),\vec{x}(\t)]$,
where $\t$ is the proper-time along the trajectory.

Following the two-level-detector paradigm, the detector's intrinsic
structure is described by two levels: the ground level
$\ket{\!\downarrow}$ and an excited one $\ket{\!\uparrow}$,
separated by an energy gap $E$ in the proper frame. It is standard
to take the interaction Hamiltonian for the model, in the
interaction picture, as

 \ba\label{Hint} H_{int}(\t) = g \: \left( e^{iE\t}
\, \ket{\!\uparrow}\bra{\downarrow\!} \: + \: e^{-iE\t} \,
\ket{\!\downarrow}\bra{\uparrow\!} \right) \;  \Phi[x^\mu(\t)] \ ,
\ea  where $g$ is the coupling constant. By this expression,
$H_{int}$ can be interpreted as a sort of boundary condition imposed
on the field along the world-line $x^\mu(\t)$. The state of the
system is the product of the field's particle content and of the
detector's state. Denoting  the field vacuum by $\vac$, the ground
level of the entire system is $\ket{0,\downarrow}$. Suppose the
system is in this state at time $\t_0$:
$\ket{\Psi(\t_0)}=\ket{0,\downarrow}$. In the absence of interaction
this is an eigenstate of the free Hamiltonian and so remain stable.
When the interaction is on, the field vacuum becomes unstable and
gets populated by quanta while the detector makes transitions.
Starting from the system vacuum at $\t_0$ , one gets, at a later
time $\t$, a system state \ba \ket{\Psi(\t)} = \hat T \: e^{\disp
-i\,\int^\t_{\t_0} \! \mbox{d}\t' \; H_{int}(\t)} \;
\ket{0,\downarrow} \ , \ea where $\hat T$ stands for time ordering.
The probability $P_+(\t)$ to find the detector in the upper level at
$\t$ (independently of the field particle content) is given by the
expectation value of the projector $\Pi_+ =
\ket{\!\uparrow}\bra{\uparrow\!}$ in the above mentioned state.
Another useful quantity--the subject of our discussion-is the
transition rate (response function) defined as $R_+(\t)\equiv
\frac{\d \disp P_+}{\d\disp\t}$. Retaining the lowest (second) order
contribution in the interaction strength $g$, we have for $\t>\t_0$
 \ba P_+(\t)
&\equiv& \scal{\Psi(\t)}{\!\uparrow}\scal{\uparrow\!}{\Psi(\t)} \label{defP+} \\
&=& g^2 \: \int^\t_{\t_0} \! \mbox{d}\t_1 \: \int^\t_{\t_0} \! \mbox{d}\t_2 \;
e^{-iE(\t_1-\t_2)} \:
W(\t_1,\t_2) \nn
&=& 2 g^2 \: \re \int^\t_{\t_0} \! \mbox{d}\t_1 \: \int^{\t_1 - \t_0}_0 \! \mbox{d}\t_2 \;
e^{-iE\t_2} \: W(\t_1,\t_1-\t_2) \label{P+}\\
R_+(\t) &=& 2 g^2 \: \re \int^{\t - \t_0}_0 \! \mbox{d}\t' \;
e^{-iE\t'} \: W(\t,\t-\t') \ ,\label{R+} \ea
 where \ba W(\t_1,\t_2)\equiv \vacave{\Phi[x^\mu(\t_1)] \, \Phi[x^\mu(\t_2)]}
\label{defWightman} \ea
 is the Wightman two-point (vacuum) function
evaluated along the trajectory, and use was made of its property:
${W(\t_1,\t_2)}^*=W(\t_2,\t_1)$. The Wightman function can be
written explicitly in terms of the trajectory:
\ba\label{defWightman2} W(\t_1,\t_2) = \inv{2{(2\pi)}^3} \:
\int\!\!\!\!\!\int\!\!\!\!\!\inte \!
\frac{\mbox{d}\vec{k}}{||\vec{k}||} \;
e^{-ik_\mu[x^\mu(\t_1)-x^\mu(\t_2)]} \ . \ea

The probability $P_-$ and the corresponding response function $R_-$
related to the de-excitation process are obtained by replacing $E$
by $-E$ in the expressions above. If a steady-state is reached, the
ratio of level populations is $n_+/n_-= R_+/R_-$.

Equations(\ref{P+}) and (\ref{R+}) exhibit causality explicitly: to
get these quantities at time $\t$ we only have to know the
trajectory prior to this time.

The quantities we deal with here enjoy an obvious and useful scaling
relation stemming from invariance under change in the units of
length: If two world-lines, $x^\mu(\t)$ and $\hat x^\mu(\hat\t)$ are
related by $\hat x^\mu(\hat\t)= \la x^\mu(\hat\t/\la)$ then their
respective response functions are related by

\ba\label{ScalingR} \hat R_+(\hat\t,E)=\la^{-1}R_+(\hat\t/\la,\la
E) \ . \ea

The intrinsic, Poincar\'{e} invariant, quantities that describe a
world-line  $x^\mu(\t)$ in $3+1$ space-time can be chosen as the
acceleration $a(\t),$ the torsion $T(\t),$ and the hyper-torsion
$H(\t),$ defined as
\ba a(\t)&=&[-a^\mu a_\mu]^{1/2} \label{defa}\\
T(\t)&=& a^{-1}[a^4 - \dot a^2 - \dot a^\mu\dot a_\mu]^{1/2} \label{defT}\\
H(\t)&=&a^{-3}T^{-2}\:\e^{\al\beta\ga\delta}\dot x_\al a_\beta \dot
a_\ga\ddot a_\delta \label{defH}\ , \ea where $\dot {} \equiv \frac{d}{d\t}$, $a^\mu=\ddot
x^\mu(\t)$, and $\e^{\al\beta\ga\delta}$ is the Levi-Civita
anti-symmetric tensor (compare Letaw \cite{Letaw}). A world-line for
which the three invariants are $\t$ independent is called
stationary. For such world-lines all points are equivalent and,
clearly, $R_+(\t)$ is constant.

The response function, being itself Poincar\'{e} invariant, can be
written as a functional of these [instead of one of the frame
specific $x^\mu(\t)$]. The above scaling relation implies for
$f(\t)=a,~H,~T$
 \ba \hat f(\hat\t)=\la^{-1}f(\hat\t/\la) \ , \ea
so that
 \ba\label{ScalingRaTH}
R_+\left[\t,E;a(\t),T(\t),H(\t)\right] = \la^{-1} \:
R_+\left[\la^{-1}\t,\la E;\la a(\la\t),\la T(\la\t),\la
H(\la\t)\right] \ . \ea (We use the convention whereby a dependent
variable is a functional of the function variables appearing after
the semicolon.) This can be used to reduce the number of variables.
For instance if we take $\la=E^{-1}$ we get \ba \label{scaleE}
R_+[\t,E;a(\t),T(\t),H(\t)] = E\: \tilde
R_+\left[E\t;\frac{a(\t_E)}{E},\frac{T(\t_E)}{E},\frac{H(\t_E)}{E}\right]
\ , \ea where $\t_E=\t/E$.

\subsection{Regularization of the Wightman function}

As expressed in \reff{defWightman2}, the Wightman function is not
well defined since the integrand behaves as $k$ times an oscillating
term in the UV limit. Various regularization procedures can be found
in the literature. For instance, one can add a small imaginary part
$-i\e$ term to $t(\t_1)-t(\t_2)$, as in Birrell and
Davies\cite{BirrellDavies}. This deformation multiplies the
integrand in \reff{defWightman2} by $e^{-k \e}$, which ensures
convergence. This gives a Wightman function of the form
\ba\label{WB&D} W^{BD}(\t_1,\t_2) = - \inv{4\pi^2} \:
\inv{[t(\t_1)-t(\t_2)-i\e]^2-[\vec{x}(\t_1)-\vec{x}(\t_2)]^2} \ .
\ea This procedure has been criticized by Takagi\cite{Takagi} and by
Schlicht\cite{Schlicht} who showed that it leads to non-physical
results even for some of the most simple examples. Takagi proposed a
regularization based on endowing the detector with a small, finite
extent, which is then described with the help of Fermi-Walker
coordinates.  Schlicht\cite{Schlicht} improved Takagi's procedure,
retaining causality explicitly in each step. This regularization is
implemented by replacing \ba\label{wfS} e^{-i k_\mu x^\mu(\t)}
\Rightarrow e^{-i k_\mu x^\mu(\t)}e^{-\e k_\mu \dot x^\mu(\t)} \ ,
\ea where $\e$ is now a measure of the proper spatial extension of
the detector. Since $k^\mu$ is light-like and $\dot x^\mu$ is future
oriented and time-like, $k_\mu \dot x^\mu(\t)>0$ ensures the
integrability for large $|\vec{k}|$. The resulting expression for
the Wightman function is

 \ba \label{WSchlicht} W^S(\t_1,\t_2) =
-\inv{4\pi^2} \: \inv{\{x^\mu(\t_1)-x^\mu(\t_2) -i\e[\dot
x^\mu(\t_1)+\dot x^\mu(\t_2)]\}^{2}} \ . \ea This procedure has been
extended to various space-time topologies\cite{Langlois} as well as
for more general kinds of spatial extents\cite{LoukoSatz}. Without
the regulator (i.e., putting $\e=0$) these two formulations of the
Wightman function are inversely proportional to the square of
geodesic distance \ba\label{GD}
d^2(\t_1,\t_2)=\eta_{\mu\nu}[x^\mu(\t_1)-x^\mu(\t_2)][x^\nu(\t_1)-x^\nu(\t_2)]
\ . \ea
\subsubsection{Proper-time regularization}

Motivated by several requirements that we deem desirable in the
Wightman function, we would like to propose another regularization.
The Wightman function always appears as a weight function in
integrations that result in observable quantities such as the
transition rate, the flux emitted\cite{PaRec}, the conditional
energy\cite{MaPa}, the flux correlation function\cite{OPa3}, \etc.
Hence, $W(\t,\t-\t')$ should be viewed in \reff{R+} as a
distribution of the running variable $\t'$. In this respect, the
analytic properties of $W$ is of key importance. More specifically,
one needs to know the poles distribution of the  Wightman function
and their relative weights in order to express analytically
quantities such as the transition rate.

As we previously noticed, apart from the coincidence point $\t'=0$
where it diverges, the Wightman function is inversely proportional
to the square of geodesic distance. This means that the analytic
properties of $W$ (its poles) come directly from the zeros of the
geodesic distance in the complex plane. We see then that the
Wightman function formally possesses only one real pole at $\t'=0$
and (possibly) an infinity of poles symmetrically distributed above
and below the real axis, since the geodesic distance is real. The
regularization is only needed to cure the divergence at $\t'=0$.
More exactly, the regularization's role is to isolate this pole such
that it does not contribute to $R_+$, only to $R_-$. As a
consequence, one finds that the difference between the excitation
and de-excitation response functions, for every trajectory, is
precisely the contribution of this pole (see next section).

From the last remarks, one deduces that it is desirable in a
regularization that it displace the $\t'=0$ pole above the real axis
such it does not contribute to $R_+$, and preserve the distribution
of other poles in. The above two procedures satisfy the first
requirement but not the second. A way to shift the zero pole while
preserving the characteristics of the world-line in a consistent way
is to introduce the regulator by replacing $x^\mu(\t)$ with
$x^\mu(\t\pm i\e)$. The wave-function in the integrand of
\reff{defWightman2} becomes \ba\label{wfeps} e^{-i k_\mu x^\mu(\t)}
\Rightarrow e^{-i k_\mu x^\mu(\t-i\e)} \ , \ea which gives the
regularized Wightman function
 \ba\label{Wieps} W^{pt}(\t_1,\t_2)
&=& - \inv{4\pi^2} \: \inv{[x^\mu(\t_1-i\e)-x^\mu(\t_2+i\e)]^2} \ .
\ea

Schlicht's Wightman function is the first-order expansion of our
\reff{Wieps} in $\e$.
 This means that the two regularizations have the same
$\e \rightarrow 0$ limit and should give the same results for
observables in this limit. The superiority of one choice over the
other may, however, show up when we want to (or have to) calculate
with a finite value of $\e$, e.g. for computational convenience.
The non-vanishing of $\e$ matters only near the coincidence point
$\t'=0$ where it appears in the denominator in
Eqs.(\ref{WSchlicht}) and (\ref{Wieps}) respectively as $-4\e^2$
and $-4\e^2+4\e^4a^2(\t)/3+\mathcal{O}(\e^6)$. Therefore,
Schlicht's Taylor expansion agrees with ours (barring the yet
higher order terms) as long as $\e\ll 1/|a(\t)|$.

Consider, for example the canonical linear motion with constant
acceleration $a_0,$ for which all the expressions can be calculated
exactly for finite $\e$. The motion is hyperbolic, with
 \ba\label{unifaccwl}
x^\mu(\t)=\left[\frac{\sinh(a_0\t)}{a_0},\frac{\cosh(a_0\t)}{a_0},0,0\right]
\ . \ea
 Schlicht's regularization gives for finite $\e$
  \ba\label{WaccS} W_{acc}^S(\t,\t-\t') = -
\frac{a_0^2}{16\pi^2} \: \left[\sinh\left(a_0\t'/2\right)-ia_0\e
\cosh\left(a_0\t'/2\right)\right]^{-2} \ , \ea whereas our proposed
proper-time regularization \reff{Wieps}, gives

\ba\label{Wacc} W_{acc}^{pt}(\t,\t-\t') = - \frac{a_0^2}{16\pi^2} \:
\sinh^{-2}(a_0(\t'-2i\e)/2) \ , \ea The corresponding transition rates
are \ba
R_+^S(\e) &=& R_+^{U} \ \times \ \frac{e^{\frac{2E}{a_0}\arctan(\e a_0)}}{1+\e^2a_0^2} \label{R+accS}\\
R_+^{pt}(\e) &=& R_+^{U} \ \times \ e^{\frac{2\pi E}{a_0}(\frac{\e a_0}{\pi}-[\frac{\e a_0}{\pi}])} \label{R+acceps}\\
R_+^{U} &=& \frac{g^2E}{2\pi}\: \inv{e^{2\pi
\frac{E}{a_0}}-1}\label{R+acc}, \ea where $R_+^{U}$ is the Unruh
\cite{Unruh} expression, which is the $\e=0$ limit of the two
regularization schemes ($[x]$ is the integer part of $x$). To
obtain our result, we have used the pole decomposition
$1/\sinh^2(x)=\sum_{n=-\infty}^{n=+\infty}1/\left(x+i n
\pi\right)^2$ and integrated by the method of residues. As
expected, expressions (\ref{R+accS}) and (\ref{R+acceps}) agree to
first order in $a_0\e$ for which they both differ from the Unruh
expression by a factor $e^{2E\e}$. So, both schemes still require
$\e E\ll1$. However, for large values of $\e a_0$ Schlicht's
formulation leads to a vanishing transition rate whereas ours
always differs from the correct expression by a finite factor
between $1$ and $e^{2\pi E/a_0}$, which is near one for $E\ll a_0$
for arbitrary values of $\e$.

For world-lines with bounded values of the acceleration we can
choose a value of $\e$ that is much smaller than $1/a$ at all times,
and then the two regularizations should give very similar results.
The situation is, however, different for world-lines with unbounded
acceleration. Consider as an example the tractable case of a linear
space trajectory with $a(\t)=-\al/\t, \; -\infty<\t<0$. For $\al\neq
1$ this is gotten, for example, for the world-line
\ba\label{trajalphasurtau}
x^\mu(\t)=\left[\half\left(-\frac{(-\t)^{1+\al}}{1+\al}-\frac{(-\t)^{1-\al}}{1-\al}\right),
\half\left(\frac{(-\t)^{1+\al}}{1+\al}-\frac{(-\t)^{1-\al}}{1-\al}\right),0,0\right];
\ea for $\al=1$ a representative world-line is

\ba \label{trajalphasurtau1} x^\mu(\t)=\left[
-\frac{\log(-\t\kappa)}{2\kappa}-\frac{\t^2\kappa}{4},
\frac{\log(-\t\kappa)}{2\kappa}-\frac{\t^2\kappa}{4},0,0 \right].
\ea

($R_+(\t)$ does not depend on $\kappa,$ as different values of
$\kappa$ are related by a Poincar\'e transformation. This trajectory
is particularly interesting because when followed by a moving mirror
it generates a constant thermal flux $\propto\kappa^2$  with
temperature $\propto\kappa$ \cite{CarlitzWilley}.)

\begin{figure}[h!]
\begin{center}
\vbox{\vss
\resizebox{7cm}{6.5cm}{\includegraphics{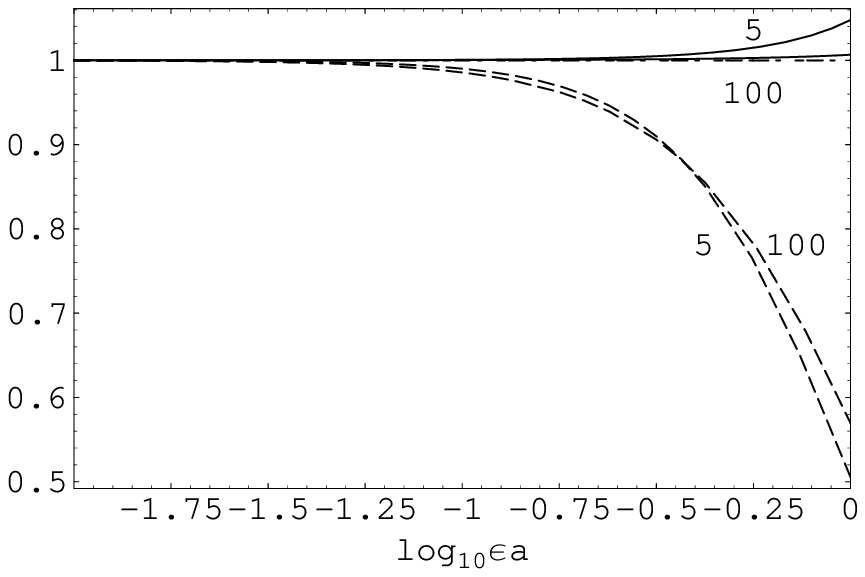}}
\hspace{5mm}
\resizebox{7cm}{6.5cm}{\includegraphics{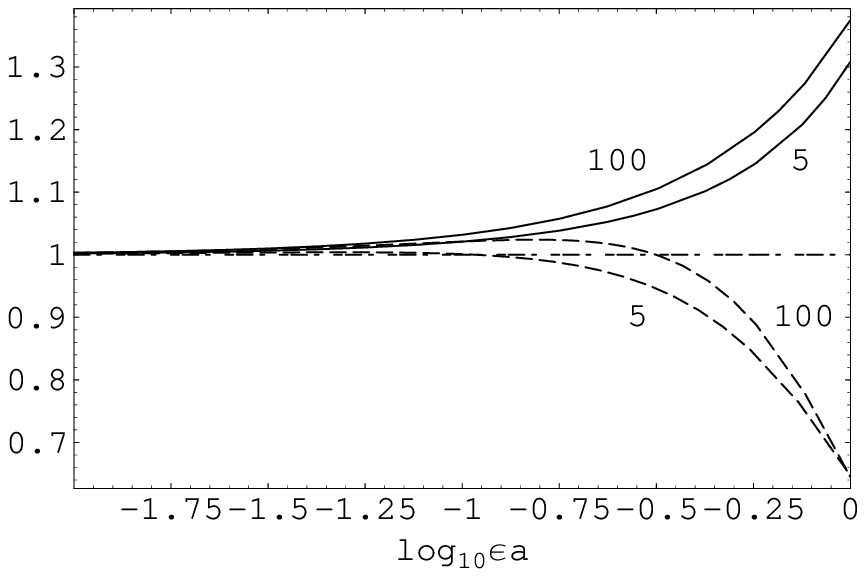}}
\vss} \caption{The ratio $R_+(\e,\t,\al,E)/R_+(0,\t,\al,E)=\tilde
R_+(a\e,\al,E/a)/R_+(0,\al,E/a)$ as a function of $a\e$ for $2\pi
E/a=10^{-2}$ (left panel) and $2\pi E/a=1$ (right panel) and for two
values of $\al=5,\ 100$, as marked. Continuous lines correspond to
the proper-time regularization, dashed curves to Schlicht's
regularization.}\label{fig1}
\end{center}
\end{figure}

For these world-lines, the scaling law, for a finite value of $\e$,
becomes (in both regularization schemes)
$R_+(\e/\la,\t/\la,\al,E\la)/\la=R_+(\e,\t,\al,E)$. Choosing
$\la=1/a$ we can write $R_+(\e,\t,\al,E)=a\tilde R_+(a\e,\al,E/a).$
The question we ask at this junction is: ``how well does the
expression for finite $\e$ approximate the correct ($\e=0$)
value?''. We present in Fig.(\ref{fig1}) the departure factor
$R_+(\e,\t,\al,E)/R_+(0,\t,\al,E)=\tilde R_+(a\e,\al,E/a)/\tilde
R_+(0,\al,E/a)$ as a function of $x\equiv a\e$, for several values
of the pair $\al,~E/a$ using Schlicht's regularization, as well as
ours. We see that, in line with what we saw in the constant
acceleration case, at least for low energies the proper-time
regularization works well for values of $\e a$ even an order larger
than for Schlicht's regularization.

\subsubsection{The subtracted Wightman function}

The troublesome singularity in the Wightman function that we wish to
regularize appears in the same way (same pole position and residuum)
in all trajectories including the inertial ones for which the
transition rate vanishes. Being time-like, all trajectories have for
small $\t$ lapses: $ d^2(\t,\t-\t')=\t'^2 + \mathcal{O}(\t'^4) $.
(The $\t'^3$ term vanishes identically for all world-lines.) So
there is some $f$ such that \ba d^2(\t,\t-\t')=\t'^2 +
\t'^4f(\t,\t-\t'), \label{exax} \ea where $f$ is regular at $\t'=0$
; in fact $f(\t,\t)=a(\t)^2/12$. We can then write using
\reff{Wieps} \ba R_+(\t) &=& -\frac{g^2}{2\pi^2}\:\re\Io{\t'}
\frac{e^{-iE\t'}}{(\t'-2i\e)^2} +\frac{g^2}{2\pi^2}\: \Io{\t'}
\frac{\cos(E\t')f(\t,\t-\t')}{1+\t'^2f(\t,\t-\t')} \ ,
\label{R+LP}\ea where we put $\e=0$ in the second term, which is now
without the pole at $\t'=0$. The first integral vanishes for
positive $E$ and $\e$, as we close the integration contour around
the lower complex plane. (In fact, this term is just the response
function for an inertial world-line.) For the de-excitation rate we
have
 \ba R_-(\t)
&=& \frac{g^2E}{2\pi} + R_+(\t). \label{R+R-} \ea
 The same pole free expression
was used in \cite{LetawPfautschCirc,Letaw} for stationary orbits,
and was generalized to non-stationary world-lines, though with some
restrictions, in \cite{LoukoSatz}, using Schlicht's regularization.
These authors  extended the applicability of the pole-free
expression to finite time interaction [Eq.(6.1) in
\cite{LoukoSatz}]. We note that this finite interaction time
expression follows straightforwardly in the proper-time
regularization scheme[Eqs.(\ref{exax})(\ref{R+LP})] applied to the
general response function \reff{R+}. It turns out that the pole-free
expression is not necessarily more useful, in numerical
determinations of the response function, than the finite $\e$
regularizations. The reason is that the function $f$ in the
integrand is defined as the difference between two diverging
expressions and its numerical evaluation requires, in the end, some
sort of regularization.

\section{The transition rate--some derived properties}

 In this section we discuss two issues regarding the response
function of general world-lines. The first is its behavior for high
detector energy gaps for stationary and non-stationary trajectories.
The second issue concerns the applicability of the
``quasi-stationary" approximation, whereby the response function is
represented by that of a stationary world-line with the
instantaneous values of acceleration, torsion, and hyper-torsion.

\subsection{High energy behavior}

To study the high energy behavior of the response function we use
the subtracted Wightman function \reff{R+LP}.  One can then write:

\ba R_+(\t) &=& \Io{\t'} \cos(E\t') F_\t(\t'), \ea where

\ba F_\t(\t') &=& \frac{g^2}{2\pi^2}
\frac{f(\t,\t-\t')}{1+\t'^2f(\t,\t-\t')}. \ea

  Following an idea in \cite{LoukoSatz}  we
integrate by parts $2N$ times to get for any $N$

\ba R_+(\t) = \sum_{n=1}^{N} \left(\frac{-1}{E^2}\right)^n
F_\t^{(2n-1)}(0) + \left(\frac{-1}{E^2}\right)^N \Io{\t'}
\cos(E\t') F_\t^{(2N)}(\t'), \label{R+LP2} \ea where we used the
fact that $F_\t^{(n)}(\t'\to\infty)=0$ since $F_\t(\t')$ decreases
at least as rapidly as $\t'^{-2}$. Using the Riemann-Lebesgue
lemma\cite{Wong}, the second term is a $\mathcal{O}(x^{-2N)})$.
So, \reff{R+LP2} can be used to pin down the asymptotic behavior
of the response function for $E\to\infty$. This we proceed to do
below, first for stationary trajectories, then for non-stationary
ones.

The first three odd derivatives of $F_\t$ are useful and are given
in terms of the world-line derivatives as:

\ba {\left(\frac{g^2}{2\pi^2}\right)}^{-1} F'_\t(0) &=& - \frac{a
\dot a}{12}, \label{fone}\\
{\left(\frac{g^2}{2\pi^2}\right)}^{-1}F_\t^{(3)}(0) &=& - \frac{\dot
a \ddot a}{12} + \frac{a^3 \dot a}{20} - \frac{a a^{(3)}}{30} +
\frac{a \dot a T^2}{60} +
\frac{a^2 T \dot T }{60}, \label{ftwo} \\
{\left(\frac{g^2}{2\pi^2}\right)}^{-1}F_\t^{(5)}(0) &=& - \frac{a^5
\dot a}{24} - \frac{a^3 \dot a T^2}{9} -  \frac{13a \dot a^3}{63} -
\frac{a^4 T \dot T}{18}\nn && + \frac{5\dot a^2 T \dot T}{42}
 +\frac{5 a \dot a \dot T^2}{42}
+\frac{53 a^2 \dot a \ddot a}{126} +\frac{5 \dot a \ddot a T^2}{84}
\nn && + \frac{5a \ddot a T \dot T}{42}
 + \frac{5a \dot a T \ddot T}{42}
+ \frac{5a^2 \dot T \ddot T}{84} +\frac{a^3 a^{(3)}}{14}\nn && +
\frac{5a a^{(3)} T^2}{252}
 - \frac{5\ddot a a^{(3)}}{42}
+ \frac{5a^2 T T^{(3)}}{252} - \frac{5\dot a a^{(4)}}{72}\nn && -
\frac{a a^{(5)}}{56} + \frac{{x^\mu}^{(4)}{x_\mu}^{(5)}}{168}.
\label{fthree}\ea

\subsubsection{Stationary trajectories}

For stationary world-lines the geodesic distance, \reff{GD}, is  a
function of the lapse $\t'=\t_1-\t_2$; so, by \reff{R+}, the
transition rate is constant (provided the interaction strength is
constant): We can write with the proper-time regularization, using
$W(\t,\t-\t')\Rightarrow W^{stat}(\t'-i\e)$, and the general
property $W^{stat}(\t')^*=W^{stat}(-\t')$
 \ba\label{R+toutbeau}
R_+^{stat}= g^2 \: \Ie{\t'} e^{-iE\t'} \: W^{stat}(\t'-i\e) \ ,
\ea

 These world-lines were classified and analyzed for
example in \cite{Letaw}. They can be grouped into six classes,
equivalent to the orbits of time-like Killing vector fields. As
said above, they are all characterized by constant values of $a,~
T,$ and $H$. For $a=T=H=0$ we get inertial world-lines; linear,
uniformly accelerated ones have $T=H=0,~a\not= 0$;  and circular,
constant speed ones have $(a<T,H=0)$ ($H=0$ for planar
trajectories).

The general property $d^2(\t_1,\t_2)=d^2(\t_2,\t_1)$ leads, for
stationary trajectories, to $F_\t(-\t')=F_\t(\t')$. Thus, all the
odd derivatives of $F_\t$ at $\t'=0$ vanish, and with them the
first sum in \reff{R+LP2}, for any $N$. We conclude then that for
all stationary trajectories $R_+$ vanishes faster than any power
of $E$ in the high energy limit. This is clearly in line with the
known thermality of the response function for uniformly
accelerated trajectories (the original Unruh case \reff{R+acc}). We discuss
other examples below.

Note first that from the general scaling law (\ref{scaleE}), we can
reduce by one the number of independent variables in the stationary
case. For example we have in this case: \ba\label{ScalingRaTHstat}
{R_+(E,a,T,H)} = E \: \tilde R_+(a/E,T/E,H/E) \ . \ea

For general planar trajectories (up to a Poincar\'e transformation)
the hyper-torsion vanishes, and the stationary trajectories are
characterized by the acceleration and the torsion. Circular
trajectories, which are defined by $T>a$, can also be defined by
some orbital radius $R$ and a frequency $\Omega$ in the frame in
which the center of the orbit is at rest, such that
$a=\ga^2\Omega^2R$ and $T=\ga a /\sqrt{\ga^2-1}$, where $\ga =
(1-R^2\Omega^2)^{-1/2}$.
 $T<a$ defines
hyperbolically unbound trajectories, and $T=a$, which is the the
 $\ga\rightarrow \infty$ limit of the circular case,
corresponds to the so called ``cusp motion''. It is realized, for
example, by the world-line
 \ba\label{Trajcusp}
x^\mu_{cusp}(\t)=\left[\t+\frac{a^2\t^3}{6}, \frac{a\t^2}{2},
\frac{a^2\t^3}{6},0\right] \ea A circular world-line is realized,
for example,  by:
 \ba\label{Trajcirc} x^\mu_{circ}(\t)=\left[\frac{T
\t}{\om}, \frac{a}{\om^2} \cos\left(\om\t\right), \frac{a}{\om^2}
\sin\left(\om\t\right),0\right], \ea where $\om\equiv
(T^2-a^2)^{1/2}$ (for circular orbits $\om=\ga\Omega$).
 The geodesic distances for these three types of world-lines is given by
 \ba
d^2(\t,\t-\t')&=&\frac{T^2}{\om^4}
\left[\om^2\t'^2-\frac{4a^2}{T^2}\sin^2\left(\om\t'/2\right)\right],
\label{GDcirc}\ea
 and is applicable also for
imaginary (hyperbolically unbound case), and vanishing (cusp case)
$\om$.

 The response function for the three stationary cases
is of the form $E$ times a function of $\nu\equiv a/E$ and
$\mu\equiv T/E$. Defining $y=E\t/2$ and
$\ka=\om/E=(\mu^2-\nu^2)^{1/2}$ we obtain with the proper-time
regularization:
 \ba
R_+ &=& \frac{g^2E}{4\pi^2} \: \Ie{y}\frac{-\ka^4 e^{-2iy}}
{\ka^2\mu^2(y-i\e)^2-\nu^2\sin^2[\ka(y-i\e)]}\ . \ea

 An analytic expression is known for the cusp motion\cite{Letaw}:
 \ba\label{R+cusp} R_+^{cusp}=\frac{g^2\: a \:
e^{-2\sqrt{3}E/a}}{8\pi\sqrt{3}} \ , \ea which, again, decreases
with $E$ faster than any power, as we found for all stationary
trajectories. This is also the response function for the circular
case in the limit of infinite $\ga$, which was derived before by
Bell and Leinaas\cite{BellLeinaas} and by
Unruh\cite{UnruhCircular}. (They proceeded by reducing the
circular geodesic distance \reff{GDcirc} to the contribution of
its first three zeros, integrating the Wightman function by the
method of residues. It can easily be understood that this gives
the same result as using the cusp geodesic distance \reff{GDcirc}
for $\om=0$.) This infinite $\ga$ ``circular Unruh effect'' is
quasi thermal in the sense that it can be described by a
``temperature'' that varies slowly with $E$, monotonically from
$\mathbf{T^<}=\frac{a}{4\sqrt{3}}$ at low energies to
$\mathbf{T^>}=\frac{a}{2\sqrt{3}}$ at high energies. Letaw and
Pfautsch\cite{LetawPfautschCirc} used the subtracted Wightman
function to numerically calculate the behavior of the transition
rate for finite $a$ and $T$. It is of interest to have an analytic
expression for $R_+^{circ}$ for finite values of $\ga$. We derive
the first order correction in $\ga^{-2}$. For the geodesic
distances one finds \ba
d^2_{circ}=d^2_{cusp}[1-\frac{a^4\t^4}{360+30a^2\t^2}
\inv{\gamma^2}+ u(a\t)\mathcal{O}(\gamma^{-4})],\ea where
$d^2_{cusp}$ is the expression for the cusp world-line
[\reff{GDcirc} with $\om=0$]; From this one deduces for fixed
$E/a$ in the limit of large $\ga$

\ba\label{gammacorrection} R_+^{circ}(E,a,\ga) =R_+^{cusp}(E,a) \:
\left[1+\inv{5\ga^2}\left(\frac{E}{a/2\sqrt{3}}-1\right) \: +
s(E/a)\mathcal{O}\left(\ga^{-4}\right)\right]\ . \ea
 We verified this with numerical results for $R_+$.
Using this and the expression for $R_-$ we see that the effective
temperature in the limit $E\rightarrow 0$ is modified to
$\mathbf{T^<_{\ga}}\approx \mathbf{T^<}(1-1/5\ga^2)$. We do not have
an analytic expression for fixed $\ga$ and high $E$.

\subsubsection{Non-stationary trajectories}

We start again with the subtracted Wightman function as in
\reff{R+LP2}.
 We noted that for a stationary trajectory all the odd
derivatives vanish because the expression for the geodesic distance
from $\t$ to $\t-\t'$ is symmetric in $\t'$ at $\t'=0$. The opposite
is also true: if there is such a symmetry for all values of $\t$ the
trajectory is stationary. Thus, if the trajectory is not stationary
there will be $\t$ values (generically, all values of $\t$) where at
least some of the odd derivatives do not vanish. For such $\t$ the
response function decays only as a power-law at high energies, and
thus, in particular, the spectrum is not thermal. For example, if
$\dot a\not =0$ the dominant high energy behavior, from \reff{fone},
is
 \ba\label{R+highenergya}
R_+(\t)&\approx& R_+^{p}(\t) = \frac{g^2}{24\pi^2}\:\frac{a\dot
a}{E^2}. \ea
 If the acceleration is constant on a world-line but the
torsion is not, the dominant contribution is, generically,
proportional to the derivative of the torsion $T(\t)$. From
\reff{ftwo} we get
 \ba R_+(\t)\approx \frac{g^2}{120\pi^2}\:\frac{a^2 T\dot T}{E^4}.
 \label{R+highenergyT}\ea
  (Note that if $a$ vanishes at
an isolated point, the first term in \reff{ftwo} gives a finite
contribution to the response function at that time.) If $T$ is also
constant the next order will dominate and it contains the derivative
of the hyper-torsion through ${x^\mu}^{(4)}{x_\mu}^{(5)}\propto H \dot H$. If this
too vanishes the trajectory is a stationary one. Conversely, if the
world-line is non-stationary, so either $a,~T,$ or $H$ do not vanish
identically, at least one of the first three orders doesn't vanish
identically. We conclude that generically the high-energy behavior
of the transition rate for non-stationary trajectories is given by
$R_+(\t)\approx g^2 \al(\t)E^{-2n}$, where $\al(\t)$ contains the
acceleration, the torsion or the hyper-torsion and their derivatives
and where $n=1,2$ or $3$.

\begin{figure}[h!]
\begin{center}
\includegraphics{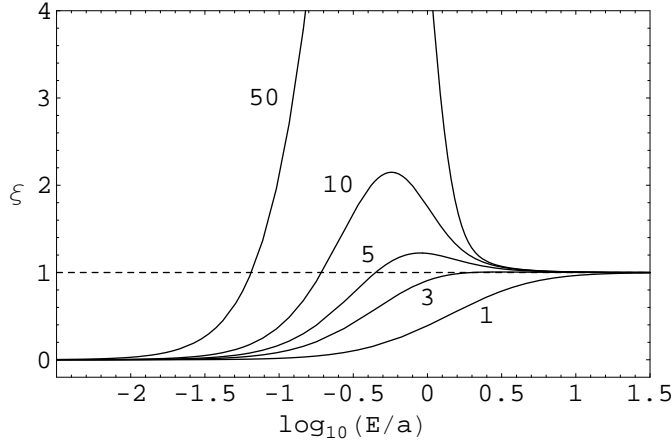}
\caption{The ratio $\xi\equiv R_+/R_+^{p}$ for the linear trajectory
\reff{trajalphasurtau} as a function of $E/a$, for different values
of $\al$ (as marked). \label{FigalphasurtauRsurRPL}}
\end{center}
\end{figure}

We checked numerically the validity of the first order approximation
for the linear trajectory
Eqs.(\ref{trajalphasurtau})(\ref{trajalphasurtau1}), having
$a(\t)=-\al/\t$. In Fig.(\ref{FigalphasurtauRsurRPL}) we plot the
ratio $\xi\equiv R_+/R_+^{p}$, calculated numerically with the
proper-time regularization. Using the scaling law we can write
$R_+(\t,\al,E)=E \tilde R_+(E/a,\al)$. Since $R_+^{p}$ is obtained
from a Taylor expansion of $R_+$, it also satisfies this scaling.
Thus, the ratio $\xi\equiv R_+/R_+^{p}$ is a function of $\al$ and
$E/a$. Fig.(\ref{FigalphasurtauRsurRPL}) shows this ratio as a
function of $E/a$ for different values of $\al$. We see that indeed
\reff{R+highenergya} is a very good approximation for the response
function at large values of $E/a$. For $\al$ larger than about 5 the
power-law behavior starts at the same value of $E/a$ regardless of
the value of $\al$. In the large-$\al$-low-$E/a$ regime $\xi$ can,
in fact, be approximated analytically very well using the
quasi-stationary approximation--see section 3.2 and
Fig.(\ref{FigalphasurtauRsurRqs}), so the behavior displayed in
Fig.(\ref{FigalphasurtauRsurRPL}) for this regime can be reproduced
analytically.

\begin{figure}[h!]
\begin{center}
\includegraphics{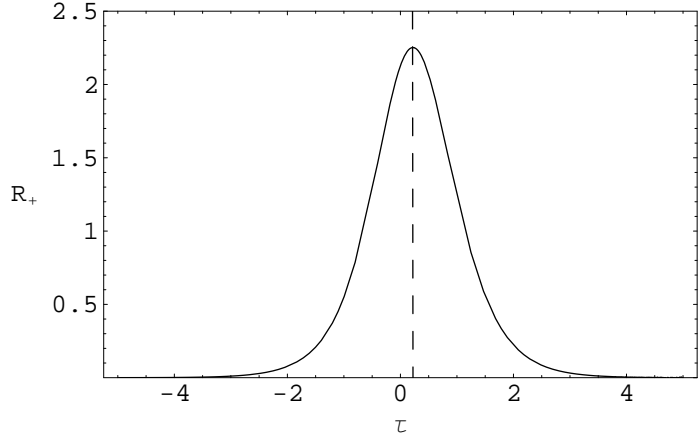}
\caption{The response function for the linear trajectory with
acceleration $a(\t)=2+\tanh\t$ for $E=20$ ($g=1$). The result is
given in units of $10^{-5}$. The dashed vertical line shows the
maximum of $R_+^{p}$, $\t_0=\ln\left[(2+\sqrt{7})/3\right]/2$.
\label{Fig2+tanhE20labarre}}
\end{center}
\end{figure}
\begin{figure}[h!]
\begin{center}
\includegraphics{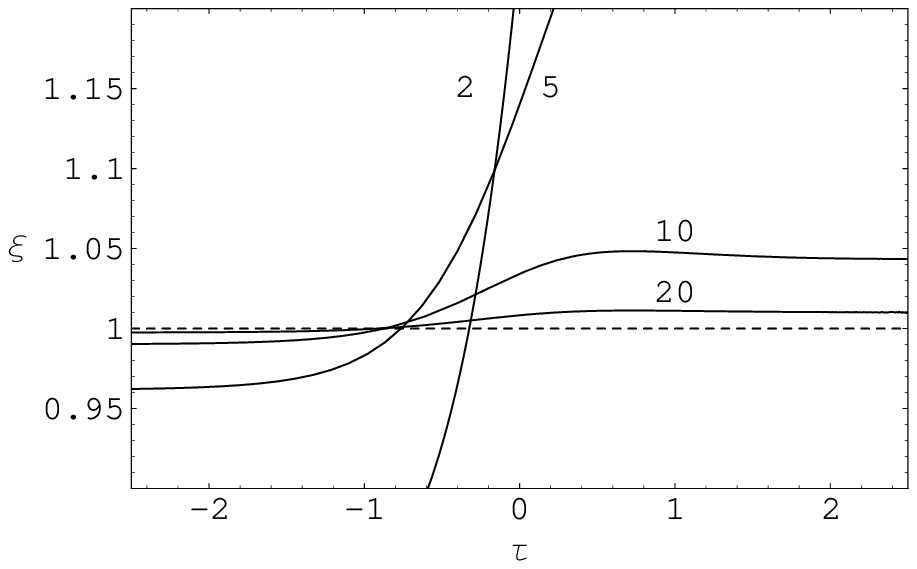}
\caption{The ratio $\xi\equiv R_+/R_+^{p}$ for the linear trajectory
\reff{traj2+tanh} with acceleration $a(\t)=2+\tanh\t$ and for
different values of the energy. \label{Fig2+tanhPL}}
\end{center}
\end{figure}

In the above example, once $E/a\gg 1$, the dominant power in $E/a$
is a good description of the response function for all values of
$\al>1$. This can be traced to the fact that in this case
 $E^{-2}F_\t^{(3)}(0)/F_\t'(0)=\mathcal{O}[(a/E)^2]$
($T\equiv 0$ for the present, linear case). In general, the
contribution of the higher orders hinges also on other
characteristics of the world-line as reflected, e.g., in the
additional terms in $F_\t^{(3)}(0)$. To better demonstrate this we
considered another example: Take a linear trajectory with
acceleration law $a(\t)=2+\tanh\t$ in some arbitrary units ($\t$ in
the inverse of the same unit); $a$ is bound between $1$ in the far
past and $3$ in the far future. Such an acceleration law is gotten
for example for the linear world-line: \ba\label{traj2+tanh}
x^\mu(\t)=\left[-e^{-\t}+\frac{e^\t}{4}+\frac{e^{3\t}}{12}+\arctan(e^{-\t}),e^{-\t}+
\frac{e^\t}{4}+\frac{e^{3\t}}{12}-\arctan(e^{-\t}),0,0\right]\ . \ea
 At large negative times the detector has a nearly constant acceleration
with the characteristic time for acceleration change
$t_{tr}=a/\dot a \simeq (3/4) e^{-2\t }\gg t_a=1/a \simeq 1$. The
first term in the expression for $F_\t^{(3)}(0)$ dominates again
and then we expect that for $E\gg 1$ the first order power-law
will be a good description. As $\t$ approaches zero the additional
terms in the expression for $F_\t^{(3)}(0)$ become important. We
first show in Fig.(\ref{Fig2+tanhE20labarre}) the computed
transition rate itself for a single value of the energy $E=20$ (in
the same units). It follows $R_+^{p}\propto \dot
a(\t)a(\t)=(2+\tanh\t)/\cosh^2\t$ rather closely (as expected
since $E\gg a(\t)$ for all $\t$). The latter is maximal at
$\t_0=\ln\left[(2+\sqrt{7})/3\right]/2$ at which it takes the
value $2(10+7\sqrt{7})/27\approx 2.11$.

   In Fig.(\ref{Fig2+tanhPL}) we plot the ratio $\xi=R_+/R_+^{p}$ for
different values of $E$. For $E$ larger than about 5 all the curves
cross a value of 1 at about the same value of $\t\simeq -0.906313$,
which is a zero of $F_\t^{(3)}(0)$.

This is a reminder that all our results are valid only to the lowest
order in $g^2$. So, for example, if we speak of asymptotic behavior it
is to be understood in the following sense: Given a high enough energy,
$E$, we assume that $g$ is small enough for the lowest order to dominate
at $E$.
We have no reason to think that our results on the asymptotic behavior
are invalidated by higher order corrections. However, even if they are, our
claim is even strengthened.
Take the non-stationary case. If higher order terms decay even faster
with $E$, then obviously they do not affect our results. If, in some case,
higher order terms decay even slower than the first order behavior we
found, then this even strengthen our conclusion regarding the non-thermal behavior
and the power-law decay.
In the stationary case where we found the first order response function
to decay faster than any power, it might in principle happen that higher
order terms decay slower.
However, this is known not to be the case at least for the Unruh case
where $R^+$  is known to all orders.

\subsection{The quasi-stationary approximation}

\begin{figure}[h!]
\begin{center}
\includegraphics{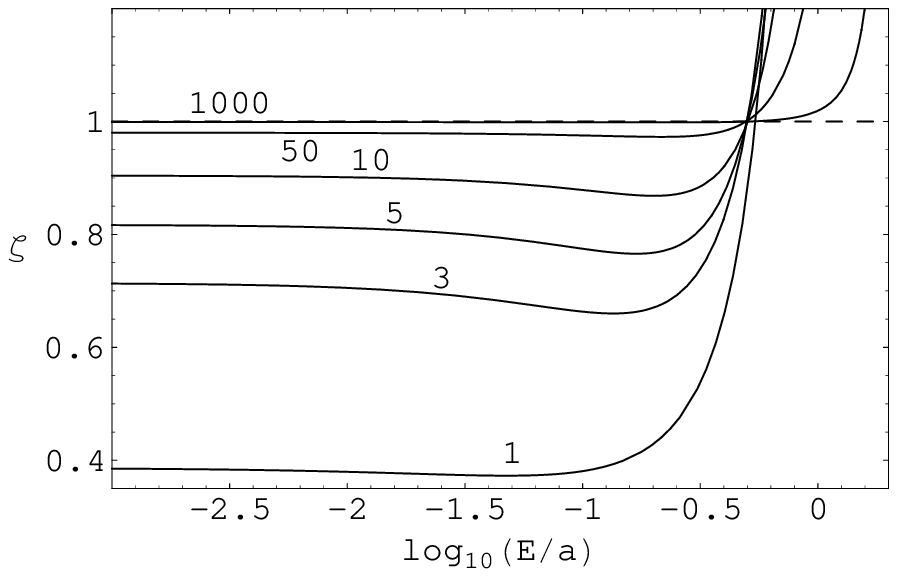}
\caption{The ratio $\zeta\equiv R_+/R_+^{q}$ as a function of
$E/a$ for different values of $\al$ (as marked) for the linear
trajectory with acceleration $a(\t)=-\al/\t$.
\label{FigalphasurtauRsurRqs}}
\end{center}
\end{figure}

One might think that if the world-line is characterized by $a,~T,$
and $H$ that vary slowly enough, the response function is well
approximated by the expression \ba \label{stat}
R_+[\t,E;a(\t),T(\t),H(\t)] \approx R_+^{q}[E,a(\t),T(\t),H(\t)]=
E\: \tilde
R_+\left[\frac{a(\t)}{E},\frac{T(\t)}{E},\frac{H(\t)}{E}\right] \
. \ea
 Here, the value of $R_+$ at $\t$, which is a functional of
$a(\t),~T(\t),H(\t)$, is approximated by its value for a
stationary world-line with the instantaneous values of these
parameters. For example,  Letaw \cite{Letaw} remarks on linear
trajectories: ``if a detector was moving with a very slowly
increasing linear acceleration one would expect the spectrum to be
Planckian, though with a slowly increasing temperature''. Our
analysis of the high energy behavior of $R_+$ shows that this
cannot be correct across the spectrum. For any given
non-stationary world-line, no matter how slow the variations in
$a,H,T$ are, the quasi-stationary approximation has to break down
at high enough energies as the spectrum becomes power-law as
opposed to that of stationary world-lines. However, at low enough
energies, slowness of variation of the three invariants might
suffice to validate the quasi-stationary approximation. It is
difficult to give general criteria for this so we treat
numerically several linear motions. In this case we write
\ba\label{R+QSlin}
R_+(\t,E) &=& R_+^{q}(\t,E) + D(\t,E) \ , \\
R_+^{q}(\t,E) &\equiv& \frac{g^2E}{2\pi}\: \inv{e^{2\pi
\frac{E}{|a(\t)|}}-1} \ ,\label{R+QSlin2} \ea where $R_+^{q}$ is
the stationary (Unruh) expression for the response function. We
wish to know under what conditions $|D(\t,E)|\ll R_+^{q}(\t,E)$.
As we said above, one necessary condition is that $E$ is small
enough, and it is convenient to express this smallness in terms of
$E/a$ (provided $a\neq 0$ locally). Another condition it that the
acceleration varies slowly enough on the world-line: the time
scale of variation of $a$: $t_{tr}\equiv a/\dot a$, has to be long
compared with $t_a\equiv 1/a$. In other words $\dot a/a^2\ll 1$.
In general, $t_{tr}$ is not the only measure of the orbital time
variations, but in the examples that we consider below it is a
good single measure. We considered, numerically again, the two
families of world-lines discussed before. In the first instance we
looked at the world-line with $a(\t)=-\al/\t$. In this case,
the time ratio  $a^2 /\dot a=\al$ is constant along the world-line, and it also represents, up to some numerical factor the time
ratio for all the possible time scales defined by the trajectory.
We show in Fig.(\ref{FigalphasurtauRsurRqs}) the ratio
$\zeta\equiv R_+/R_+^{q}$ as a function of $E/a$ for several
values of the parameter $\al$. Because of the scaling properties
mentioned above, $\zeta$ is a function of only these two
variables, and is given by: \ba \zeta = \frac{e^{2 \pi
E/a}-1}{\pi\alpha E/a} \: \Io{\t'} \cos\left(\alpha \t' E/a\right)
\: \left\{
\frac{\alpha^2-1}{\left[1-(1+\t')^{1-\alpha}\right]\left[1-(1+\t')^{1+\alpha}\right]}
+ \inv{\t'^2} \right\}. \ea

We see that indeed the quasi-stationary approximation is good
($\zeta\approx 1$) when $E/a$ is small and the orbital variations
are slow (high $\al$), but it breaks down for all values of $\al$
plotted for large $E/a$, and for all values of $E$ if $\al$ is
small. The parameter $\zeta$ has a well defined low energy limit,
which for high $\al$ behaves, according to the numerical results, as
$\zeta(E=0,\al)\approx 1-\al^{-1}$.

\begin{figure}[h!]
\begin{center}
\includegraphics{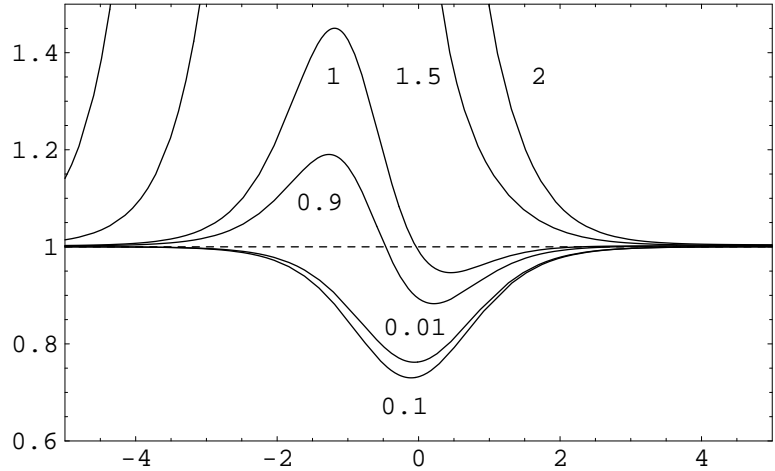}
\caption{The ratio $\zeta\equiv R_+/R_+^{q}$ for different values of
the energy for the linear trajectory with $a(\t)=2+\tanh\t$.
\label{Fig2+tanhRsurRqs}}
\end{center}
\end{figure}
Next, we reconsidered the linear trajectory \reff{traj2+tanh} with
acceleration $a(\t)=2+\tanh\t$. Since the acceleration is bounded
between 1 and 3, the condition $E\ll a$ is tantamount to $E\ll 1$.
The ratio of time scales $\dot a/a^2$ is small for large, positive
and negative, values of $\t$ where it decreases exponentially, but
is of order one near $\t=0$.
In Fig.(\ref{Fig2+tanhRsurRqs}) we plot $\zeta$ as a function of
$\t$ for different values of $E$. For $E=2$--which corresponds to
$E/a\approx 2$ for large negative $\t$, and $E/a\approx 2/3$ for
high positive $\t$--we see that $\zeta$ departs from one appreciably
even when the acceleration varies very slowly. Also for very small
values of $E$ the quasi-stationary approximation is valid only when
the acceleration varies on time scales longer than $1/a$ (i.e., away
from $\t=0$).

Finally, we consider a linear trajectory with an acceleration law
$a(\t)=\t_0^{-2/3}(-\t)^{-1/3}$  ($\t<0$). It is realized, e.g.,
by the world-line \ba\label{trajBH} x^\mu(\t)=\left[ \t_0
\sqrt{\frac{2}{3}} \left( \Iinfsup{0}{A}{x} \sinh x^2 - A \sinh
A^2 \right) , -\t_0 \sqrt{\frac{2}{3}} \left( \Iinfsup{0}{A}{x}
\cosh x^2 - A \cosh A^2 \right) , 0,0\right] \  , \ea where
$A=\sqrt{\frac{3}{2}} \left(\frac{-\t}{\t_0}\right)^{1/3}$. This
acceleration law might be of interest because it reproduces the
time dependence of the surface acceleration of an evaporating
black hole (in which case $\t$ is the proper-time of a static
observer at infinity). Accelerating detectors and black hole are
related (e.g., \cite{Unruh}): An analogy can be drawn between the
Unruh thermal bath for a uniform acceleration $a$, and Hawking
radiation from a constant mass black hole with surface gravity
$\kappa=1/4 M$, as the temperatures for the two cases are given
respectively by
  \ba k_BT_U=\frac{\hbar a}{2\pi c} \: ,
\: k_BT_H=\frac{\hbar\kappa}{2\pi c} \ . \ea If we apply the
semiclassical treatment down to complete evaporation through a massless scalar field,
and take the time $\t=0$ at this event, one finds that $\kappa = T_P^{-2/3}
(-\t)^{-1/3},$ where $T_P=t_{Pl}/\sqrt{80\pi}$ and $t_{Pl}$ is the Planck time
(the numerical factor relating $T_P$ and $t_{Pl}$ varies when considering other fields
according to their mass, spin, \etc).

In discussing black hole evaporation one usually assumes the
analogue of our quasi-stationary approximation; i.e., that at any
given time a black hole radiates as a black body corresponding to
its momentary mass. We set to check to what extent  such an
approximation is justified in our mock-evaporation case\footnote{The
process of Hawking radiation has been proven to keep its thermal
character until the mass of the black-hole reaches $\mathcal{O}(M_{Pl})$
(see \cite{Bardeen} for a heuristic picture, \cite{PaPiran} for a
direct numerical treatment and \cite{Massar} for a semiclassical
back-reaction analysis). Our analogy is an alternative, heuristic
check.}. Of course, our chosen acceleration law is based on the
quasi-stationary approximation in deducing the time dependence of
the mass. It is also clear that the analogy becomes meaningless
beyond about $\t=-T_P$. But we still consider the analogy at all
times as it may be of interest for its own sake: in general our
parameter $\t_0$ need not be related to the Planck time. Again, we
consider the parameter $\zeta=R_+/R_+^{q}$ as a measure of the
departure from quasi-stationarity. The scaling law for $R_+$ tells
us that this quantity can be written as a function of two
parameters; for example $\zeta=\zeta(\t/\t_0,E/a)$, or alternatively
$\zeta=\tilde\zeta(\t/\t_0,E\t_0)$. The first helps to follow the
time dependence of $R_+$ at a fixed value of $E/a$, which tells us
where on the black body spectrum we are; the second representation follows the time evolution for a
fixed value of the energy. The time variable $-\t/\t_0 =(3 \dot a /a^2)^{-3/2}$
is also the parameter we use to measures the fastness of the variation of $a$.

\begin{figure}[h!]
\begin{center}
\includegraphics{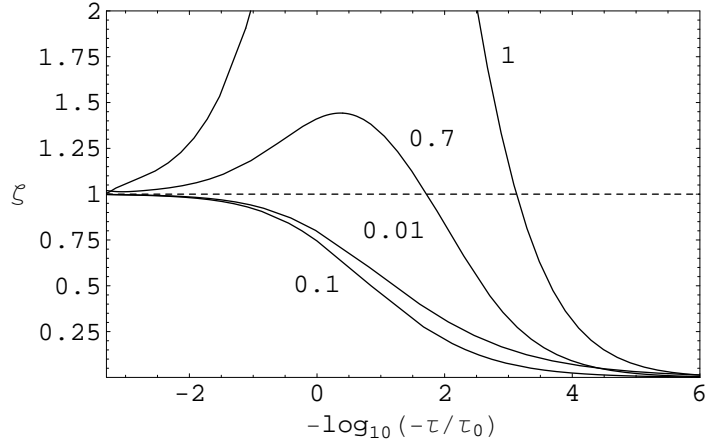}
\caption{The ratio $\zeta=R_+/R_+^{q}$
as a function of $\t/\t_0$
for different values of $E/a$ (as marked)
for the linear trajectory with acceleration $a(\t)=\t_0^{-2/3}(-\t)^{-1/3}$. \label{FigBHEsura}}
\end{center}
\end{figure}
\begin{figure}[h!]
\begin{center}
\includegraphics{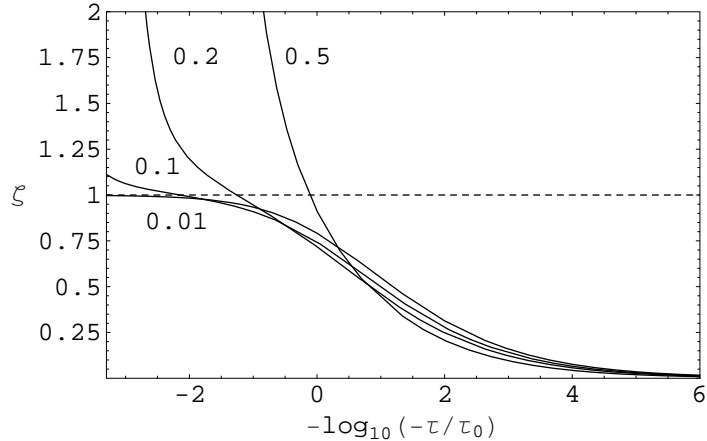}
\caption{The ratio $\zeta=R_+/R_+^{q}$
as a function of $\t/\t_0$
for different values of $E\t_0$ (as marked)
for the linear trajectory with acceleration $a(\t)=\t_0^{-2/3}(-\t)^{-1/3}$. \label{FigBH}}
\end{center}
\end{figure}
We plot the above two representations of $\zeta$ (computed with the proper-time
regularization) in Figs.(\ref{FigBHEsura}) and (\ref{FigBH}) respectively.
In the first $\zeta$ is plotted vs. $\t/\t_0$
for different values of $E/a$ and in the second for different values of $E\t_0$.
(For values of these parameters below the minimum ones plotted the curves remain the same.)
We see that, in line with our previous
results, for high enough energies, or rapid enough variations in the acceleration,
the quasi-stationary approximation breaks down appreciably.
For $-\t/\t_0>100,$ where
$\dot a/a^2$ is about 0.015, the quasi-stationary approximation is valid for all energies around and below the
black body peak at $E/a=1/2\pi$. In the black hole analogy this value of $\t/\t_0$ would correspond to a
black hole mass of $\approx0.07 M_{Pl}$ ($M_{Pl}$ is the Planck mass). So our results would indicate that
the quasi-stationary approximation for black holes holds in the whole classical regime.

 One can show
that, in the limit $\t\to 0$, $R_+$ tends to an energy-dependent
finite value $R_+^0(E)$ (unlike the $a(\t)=-\alpha/\t$ case, where
$R_+$ diverges in this limit). We found numerically that
$R_+^0(E)\stackrel{E\to 0}{\longrightarrow} (2/15\pi)(g^2/\t_0)$.

We see then that at least for the above three examples $\dot
a/a^2\ll 1$ is indeed a sufficient condition for the validity of the
quasi-stationary approximation, provided $E/a\lesssim 1$.

\section{Conclusion}

We considered several issues concerning the behavior of the
transition rate for a moving two-level detector in the vacuum. We
introduced the Feynmann prescription for regularizing the Wightman
function in order to secure causality for the response function of
a two-level detector. We then use this to derive certain
properties of the response function. We showed that for stationary
world-lines the response function decreases faster than any
power-law in the high energy limit, in concordance with the known
exponential decrease for a couple of known special cases. In
contrast, non-stationary motions are characterized by a power-law
decline of the response function at high detector energies:
Generically, the response function at any given time vanishes as
$E^{-2}$ at high energies; in special cases the decline is faster:
either as  $E^{-4}$ or as $E^{-6}$. We also considered some
aspects of the applicability of the quasi-stationary
approximation, whereby the response function is approximated by it
value for a stationary trajectory with the instantaneous values of
the invariant parameters $a, ~T, ~H$. Our result on the high
energy behavior precludes this approximation for high energies
(generically $1\lesssim E/a$), but in the low energy regime we
find for several examples studied numerically that the
approximation is good when $\dot a/a^2\ll 1$.


${\bf{Acknowledgements}}$
\newline
We thank R. Parentani and T. Polack for useful discussions.

\end{document}